\documentclass{ws-ijmpb}
\usepackage{graphicx}
\begin{document}

\title{Time-dependent current-density functional theory
for the friction of ions in an interacting electron gas }

\author{V.~U.~Nazarov}

\address{Research Center for Applied Sciences, Academia Sinica,
 Taipei 115, Taiwan}

\author{J.~M.~Pitarke}
\address {CIC nanoGUNE Consolider, Mikeletegi Pasealekua 56, E-2009 Donostia, Basque Country}
\address {Materia Kondentsatuaren Fisika Saila, UPV/EHU and Unidad Fisica de Materiales,
CSIC-UPV/EHU, 644 Posta kutxatila, E-48080 Bilbo, Basque Country}

\author{Y.~Takada}
\address {Institute for Solid State Physics, University of
Tokyo, Kashiwa, Chiba 277-8581, Japan}

\author{G.~Vignale}
\address {Department of Physics and Astronomy, University of Missouri, Columbia, Missouri 65211, USA}

\author{Y.-C.~Chang}
\address{Research Center for Applied Sciences, Academia Sinica,
 Taipei 115, Taiwan}

\maketitle

\begin{abstract}
Due to the strongly nonlocal nature of $f_{xc}({\bf r},{\bf r}',\omega)$ the {\em scalar}
exchange and correlation (xc) kernel of the time-dependent density-functional theory (TDDFT),
the formula for Q the friction coefficient of an interacting electron gas (EG) for ions tends
to give a too large value of Q for heavy ions in the medium- and low-density EG,
if we adopt the local-density approximation (LDA) to $f_{xc}({\bf r},{\bf r}',\omega)$,
even though the formula itself is formally exact.
We have rectified this unfavorable feature by reformulating the formula for Q in
terms of the {\em tensorial} xc kernel of the time dependent current-density functional theory,
to which the LDA can be applied without intrinsic difficulty.
Our numerical results find themselves in a considerably better agreement
with the experimental stopping power of Al and Au for slow ions than those previously obtained within the LDA to the TDDFT.
\end{abstract}

%\pacs{71.15.Mb; 34.50.Bw}

\section{Introduction}
The problem of the stopping power (SP)
of solids for moving ions (the energy-loss by ion per its unit path-length)
has been attracting much attention for decades
(see, e.g., Ref.~\cite{Sabin-04} for recent reviews).
Theoretically, this problem has been investigated in two distinct ways:
One is a perturbation approach in which the projectile-target interaction is taken as an expansion parameter \cite{Echenique-90}.
The other is based on the density-functional theory (DFT), treating the problem in a fully nonlinear manner from
the outset \cite{Echenique-81,Echenique-86}.

For slow projectiles, the perturbation expansion generally fails \cite{Nazarov-04}, making the
nonlinear treatment indispensable.
In the limit of zero velocity, the potential scattering (binary-collision)
theory is known to give an expression for the friction coefficient (FC)
(SP divided by the projectile velocity at its zero value) of electron gas (EG)
as \cite{Finneman-68,Echenique-81,Echenique-86}

\begin{equation}
Q= \bar n_0  \, k_F \sigma_{tr}(k_F),
\label{start1}
\end{equation}
where $k_F$ is the Fermi momentum,
$\sigma_{tr}(k_F)$
is the transport cross-section of the elastic scattering of an electron at the Fermi level
in the self-consistent Kohn-Sham  (KS) \cite{Kohn-65}
potential $V_{KS}(r)$ of the ion statically immersed in the EG,
and $\bar n_0$ is the density of the homogeneous EG in the absence of the ion.
Equation (\ref{start1}) is, however, incomplete \cite{Nagy2-89,Sayasov-93} in the sense that
it ignores the role of the {\em dynamical exchange and correlations} (xc) effects
in the problem of the ion slowing.
These effects cannot be neglected even at vanishingly small velocities.
The complete formal solution to the problem of the friction coefficient
of the interacting EG including the dynamical xc effects has been recently found as \cite{Nazarov-05}
\begin{equation}\label{s}
Q = Q_1 + Q_2,
\label{Q12}
\end{equation}
where
\begin{eqnarray}
Q_1&=& - \int
\frac{[\nabla_{\bf r} V_{KS}(r)\cdot {\bf v}]}{v}  \frac{[\nabla_{{\bf r}'} V_{KS}(r')\cdot {\bf v}]}{v}\cr\cr
&\times& \frac{\partial {\rm Im} \chi_{KS}({\bf r},{\bf r}',\omega)}{\partial\omega}
\Bigr |_{\omega=0} d{\bf r}  \, d{\bf r}',
\label{Q1}\\
Q_2 &=& - \int
\frac{[\nabla_{\bf r} n_0(r)\cdot {\bf v}]}{v}  \frac{[\nabla_{{\bf r}'} n_0(r')\cdot {\bf v}]}{v}\cr\cr
&\times& \frac{\partial {\rm Im} f_{xc}({\bf r},{\bf r}',\omega)}{\partial\omega}
\Bigr |_{\omega=0} d{\bf r} \, d{\bf r}'.
\label{Q2}
\end{eqnarray}
In Eqs. (\ref{Q12})-(\ref{Q2}),  $Q_1$ and $Q_2$ denote, respectively, the independent-electrons and the dynamical xc
contributions to the friction coefficient; $\chi_{KS}({\bf r},{\bf r}',\omega)$ and
$f_{xc}({\bf r},{\bf r}',\omega)$ are the KS density-response function and the dynamical
xc kernel \cite{Gross-85}, respectively, of the inhomogeneous
many-body system of an ion at rest in the EG, $n_0(r)$ is the ground-state density
of this system, and ${\bf v}$ is a small velocity of the projectile, on the value of which the final results evidently
do not depend.

The dynamical xc kernel is defined as the Fourier transform with respect to time
of the functional derivative of the time-dependent xc potential with respect to the time-dependent
electron density
\begin{equation}\label{fxc}
f_{xc}[n_0({\bf r})]({\bf r},t;{\bf r}',t')=\left.{\delta V_{xc}[n]({\bf r},t)\over\delta n({\bf r}',t')}\right|_{n=n_0({\bf r})}.
\end{equation}
A formal proof of Eqs. (\ref{Q12})-(\ref{Q2}) is provided in Appendix \ref{Pr1}.

For the independent-particle contribution,
Eq.~(\ref{Q1}) can be shown to coincide exactly with
Eq.~(\ref{start1}) of the binary-collision theory (see Appendix \ref{Q1start1} of the present paper).
Actual evaluation of Eq.~(\ref{start1}) can be done through a well known
procedure \cite{Echenique-86} of the self-consistent solution of the KS equations
\begin{eqnarray*}
\left[- (1/2) \, \Delta + V_{KS}(r)\right] \psi_{i}({\bf r})
= \epsilon_i \psi_{i}({\bf r}),
\end{eqnarray*}
where $\psi_{i}({\bf r})$ and $\epsilon_i$ are the single-particle wave-functions and the eigenenergies,
respectively, in the potential
\begin{eqnarray}
V_{KS}(r)&=& - Z_1/r + \int  [n_0(r')-\bar n_0]/|{\bf r}-{\bf r'}| \, d{\bf r}'
\label{veff}\\
&+&V_{xc}[n_0(r)]-V_{xc}(\bar n_0),
\nonumber
\end{eqnarray}
where $Z_1$ is the atomic number of the ion, and $V_{xc}[n_0(r)]$ is the xc potential of the
static DFT. The ground-state density is determined as
\begin{equation}
n_0(r)= \sum\limits_{\epsilon_i\le \epsilon_F} |\psi_{i}({\bf r})|^2,
\label{n0}
\end{equation}
where $\epsilon_F$ is the Fermi energy. Upon achievement of self-consistency,
the transport cross-section  can be conveniently found through \cite{Huang-48}
\begin{eqnarray}
\sigma_{tr}(k)=\frac{4\pi}{k^2} \sum\limits_{l=0}^{\infty} (l+1) \sin^2[\delta_l(k)-\delta_{l+1}(k)],
\label{trans}
\end{eqnarray}
where $\delta_l(k)$ is the phase-shift of the scattering state with  momentum
$k$ and  angular momentum $l$.

Determination of both the noninteracting-electron ($Q_1$) and the dynamical xc ($Q_2$)
parts of the friction coefficient require, respectively, pertinent approximations to
the static xc potential $V_{xc}(r)$ and the dynamical xc kernel $f_{xc}({\bf r},{\bf r}',\omega)$.
The local-density approximation (LDA)  provides  simple and rather accurate
schemes \cite{Gunnarsson-76,Perdew-81} for the quantities derived from the static xc potential.
The situation with the dynamical xc kernel is, however, much more complicated:
The scalar xc kernel $f_{xc}({\bf r},{\bf r}',\omega)$ of the time-dependent density functional theory (TDDFT)
is known to be a strongly nonlocal function of space coordinates \cite{Vignale-95},
implying a difficulty in directly constructing an accurate approximation  to calculate $Q_2$ through Eq. (\ref{Q2}).
On the other hand, it is known \cite{Vignale-96}
that if the current-density rather than the particle-density
is chosen as the basic variable of the theory, then the corresponding
tensorial xc kernel is a function of short range.
This fact made it possible to construct a consistent LDA scheme
within the framework of the time-dependent current-density functional theory \cite{Vignale-96,Vignale-97}
(TDCDFT).
More recently, the TDCDFT has been successfully applied in
studying the damping of excitations in disordered systems \cite{Ullrich-02}
and dynamical corrections to the resistance of  point contacts \cite{Sai-05}.

Therefore the main purpose of this paper is to recast Eq. (\ref{Q2}) into a form expressed
through the tensorial xc kernel of the TDCDFT which allows the LDA treatment.
With using this form, we have made comparison between TDDFT and TDCDFT
through numerical investigation to find that $Q_2$ is overestimated in
the LDA to the TDDFT for heavy ions immersed in the medium- and
low-density EG, while the LDA to the TDCDFT rectifies this unfavorable feature.

It must be noted, that the {\em exact} TDDFT and the {\em exact} TDCDFT,
were they possible to develop,
would yield, of course, exactly the same friction coefficient $Q$.
Both theories would then involve {\em nonlocal} scalar and tensorial
xc kernels, respectively. The basic idea of our method is to express
quite generally
the scalar xc kernel of TDDFT through the tensorial one of TDCDFT [Eq. (\ref{via})].
Then we show that if the local tensorial xc kernel of TDCDFT is used
on the right-hand side of this expression, the resulting {\em nonlocal} scalar xc kernel
of TDDFT is free of the inconsistencies of the LDA to TDDFT, and when used in
Eq (\ref{Q2}), it yields a good description of the many-body dynamical
effects in the friction coefficient of the interacting EG for ions.

In Refs.~\cite{Nazarov-05,Nazarov-07} we have summarized main results
of the TDDFT and TDCDFT approaches, respectively, to the problem of the stopping power of metals for slow ions.
The present work gives rigorous proofs of all the assertions  made in Ref.~\cite{Nazarov-05,Nazarov-07}
as well as specifies details of both the mathematical formalisms and the computational procedures.

The organization of this paper is as follows: In Sec.~\ref{TDDFT} and
Appendices A - \ref{Q1start1} we present a formal TDDFT of the friction of ions in EG
and in Subsection \ref{LDAS} we discuss a fundamental difficulty  the LDA to the dynamical
xc kernel encounters, i.e.,
the violation within LDA of the requirement
that the friction coefficient of free space
for isolated atoms and ions must be zero.
In Sec.~\ref{CDFT} we develop the TDCDFT of the friction of ions in EG
and demonstrate the resolution within the LDA to the TDCDFT of the above contradiction
(Subsection~\ref{RVP} and Appendix~\ref{srVproof}). In Sec.~\ref{Num}
we outline the calculational procedures, present the results and their discussion.
Our conclusions are collected in Sec.~\ref{Conc}.
Appendix \ref{Expl} provides explicit expressions for the response functions
and xc kernel in the spherical coordinate system that we use in our calculations.

\section{Time-dependent density-functional theory of the friction coefficient}
\label{TDDFT}
We consider a recoiless probe particle of the charge $Z_1e$, where $e$ is the absolute value of the charge of electron,
moving with
velocity ${\bf v}$ in the otherwise homogeneous gas of interacting
electrons at zero temperature. The stopping power
is the retarding force that the polarization
charge distribution in the vicinity of the projectile exerts on the
projectile itself. Accordingly, one can write
\cite{Echenique-90} (we use atomic units throughout the paper)
\begin{equation}\label{0}
-\frac{dE}{dx}
= - \frac{Z_1}{v}\!\int \! d{\bf r}\,
d{\bf r}'\,\delta({\bf r}\!-\!{\bf v}t)\,
{\bf v}\!\cdot\!{\bf\nabla}_{\bf r}\,n_{ind}({\bf r}',t)
/|{\bf r}\!-\!{\bf r}'|,
\end{equation}
$n_{ind}$ being the electron density induced by the projectile.
Our first step is  to obtain the following expression for the friction coefficient
of the {\em interacting} EG:
\begin{eqnarray}
&&Q \equiv \lim_{v\rightarrow 0} - \frac{1}{v} \frac{dE}{dx} =
\label{Qint}\\
&&- \int
\frac{[\nabla_{\bf r} V_0(r)\cdot {\bf v}]}{v}  \frac{[\nabla_{{\bf r}'} V_0(r')\cdot {\bf v}]}{v}
\frac{\partial \,{\rm Im} \chi({\bf r},{\bf r}',\omega)}{\partial\omega}
\Bigr |_{\omega=0} d{\bf r}  \, d{\bf r}' \nonumber,
\end{eqnarray}
where $V_0(r)= - Z_1/r$ is the
bare Coulomb potential and $\chi({\bf r},{\bf r}',\omega)$
is the linear density-response function
of the system of {\it interacting} electrons in the static field of
an impurity of charge $Z_1$ at ${\bf r}=0$.
A proof of Eq.~(\ref{Qint})
is provided in Appendix
%\ref{P1}
A.
We emphasize that Eq.~(\ref{Qint}) is fully nonlinear with respect to
the interaction of the bare charge $Z_1$ with the EG.
The next step, which is performed in Appendix \ref{Pr1}, is to show that Eq.~(\ref{Qint})
leads to Eqs.~(\ref{Q12}) - (\ref{Q2}). The equivalence between Eq.~(\ref{Q1}) and the friction coefficient of Eq.~(\ref{start1})
from the binary-collision theory   is proven in Appendix \ref{Q1start1}.

\subsection{LDA to the TDDFT and
the violation of the requirement  for the friction coefficient of free space to be zero}
\label{LDAS}
Within the LDA, the scalar dynamical xc kernel is given by \cite{Gross-85}
\begin{equation}\label{fxc0}
f_{xc}({\bf r},{\bf r}',\omega)=f^h_{xc,L}[n_0(r),\omega]\,\delta({\bf
r}-{\bf r}'),
\end{equation}
where $f^h_{xc,L}(n,\omega)$ denotes the $q\to 0$ limit of the
frequency-dependent longitudinal xc kernel of a homogeneous EG of
the density $n$. Due to the spherical symmetry of $n_0(r)$,
substitution of Eq.~(\ref{fxc0}) into Eq.~(\ref{Q2}) yields
\begin{equation}
Q_2=
-\frac{4\pi}{3}  \int\limits_0^\infty
dr\left[r\,n_0'(r)\right]^2
\frac{\partial {\rm Im}f^h_{xc,L}[n_0(r),\omega]}
{\partial\omega}\Bigr |_{\omega=0}.
\label{fxcl}
\end{equation}

An important test of the theory is the limit of zero density of the EG, i.e.,
$\bar n_0\rightarrow 0$.
It is evident that the friction coefficient for any atom (ion) should be zero in this case.
The noninteracting-electrons part $Q_1$ of Eq.~(\ref{start1}) obviously satisfies this requirement.
However, it is  easy to see from Eq.~(\ref{fxcl}) that within the LDA, $Q_2$
remains finite since the ground-state density $n_0(r)$ of an isolated ion (atom)
has a nonzero gradient, and the frequency derivative of ${\rm Im}f^h_{xc,L}(n,\omega)$ is always nonzero
and negative \cite{Qian-02}. Therefore, the LDA to the scalar xc kernel
does not satisfy the requirement of the friction coefficient
of vacuum for an atom to be zero.

Let us now demonstrate that the rigorous formula (\ref{Q2}) does pass the above test.
For an arbitrary bounded (finite) system,
which is an atom in vacuum,
a sum rule holds \cite{Vignale-95}
\begin{equation}
\int f_{xc}({\bf r},{\bf r}',\omega) \, \nabla_{r'} n_0({\bf r}') \, d {\bf r}' = \nabla_{r} V_{xc}({\bf r}).
\label{srV}
\end{equation}
Substituting Eq.~(\ref{srV}) into Eq.~(\ref{Q2}) and noting that the right-hand side of the former
does not depend on frequency, we obtain zero identically, which proves our assertion.
We note that the sum rule (\ref{srV}) {\em does not} hold for extended systems
and, therefore, $Q_2$ is finite for atoms (ions) in EG of nonzero density.

\begin{table}[h]
\tbl{Inaccuracy of LDA to  TDDFT: Friction coefficient of free space
  ($r_s =  \infty$) and that of an electron liquid of $r_s  =  2.2$  for several atoms.
  Line 3 is the ratio of lines 1 and 2 (\%).}
{\begin{tabular}{c c c c c c c c}
  \hline
    Atom  & He & Be & C & O & Ne & Mg & Si \\
  \hline
  Q($r_s =  \infty$) & 0.04 & 0.11  & 0.17 & 0.24 & 0.30 & 0.36 & 0.43\\
  Q($r_s$=2.2) & 0.34 & 0.43  & 0.70 & 0.46 & 0.16 & 0.15 & 0.54 \\
  \%             & 12 & 25 & 24 & 52 & 188 & 240 & 80 \\
    \hline
\end{tabular}}
\end{table}

To give an idea how large can be the error introduced by LDA to TDDFT,
in Table 1 we list the values of FC of free space ($r_s=\infty$) for a number of atoms.
For comparison, in column 3 of this table we list the FC of EG of $r_s=$2.2 for
the same atoms obtained with use of Eq. (\ref{start1}) (neglect of dynamical xc).

The shortcoming of the LDA to the TDDFT manifested in the the violation
of the requirement
of zero friction coefficient of vacuum
will be overcome in the next section within the LDA to the TDCDFT.

\section{Time dependent current-density functional theory of the friction coefficient}

\label{CDFT}
The purpose of this section is to express Eq.~(\ref{Q2}) for the dynamical xc
contribution to the friction coefficient in terms of the xc tensor of the time-dependent TDCDFT
in order to overcome
the difficulties the LDA encounters when applied to the ordinary TDDFT as discussed in the previous section.
We know that \cite{Vignale-96}
\begin{equation}
\hat{f}_{xc,ij}({\bf r},{\bf r}',{\omega})=
\hat{\chi}^{-1}_{KS,ij}({\bf r},{\bf r}',{\omega})-
\hat{\chi}^{-1}_{ij}({\bf r},{\bf r}',{\omega})+ \frac{c}{\omega^2} \nabla_i \frac{1}{|{\bf r}  - {\bf r}'|}\, \nabla'_j,
\label{ft}
\end{equation}
where $\hat{f}_{xc,ij}$ is the tensorial xc kernel of the TDCDFT, $\hat{\chi}_{ij}$ is the many-body current density-vector potential
response function, and $\hat{\chi}_{KS,ij}$ is the single-particle KS counterpart of the latter.
On the other hand, the scalar xc kernel of the ordinary TDDFT can be written as
\begin{eqnarray}
f_{xc}({\bf r},{\bf r}',{\omega})= \chi^{-1}_{KS}({\bf r},{\bf
r}',{\omega})\! - \! \chi^{-1}({\bf r},{\bf r}',{\omega})\! - \!
\frac{1}{|{\bf r}-{\bf r}'|}, \label{fs}
\end{eqnarray}
where $\chi$ is the longitudinal density-scalar potential response function and $\chi_{KS}$ is its
Kohn-Sham counterpart. We can write in operator notations
\begin{eqnarray}
\chi= -\frac{c}{\omega^2}
\nabla \cdot \hat{\chi} \cdot \nabla,
\label{chit}
\end{eqnarray}
and consequently
\begin{eqnarray}
\chi^{-1} = -\frac{\omega^2}{c}
\nabla^{-2} \nabla \cdot \left(\hat{L} \hat{\chi} \hat{L} \right)^{-1} \cdot
\nabla  \nabla^{-2},
\label{chi1}
\end{eqnarray}
where $\hat{L}$ is the longitudinal projector operator
\begin{eqnarray*}
\hat{L}_{ij}=\nabla_i \nabla_j \nabla^{-2}.
\label{LP}
\end{eqnarray*}

Using a simple operator identity
\begin{eqnarray}
\left(\hat{L} \hat{\chi} \hat{L} \right)^{-1} =
\hat{L} \hat{\chi}^{-1} \hat{L} -\hat{L} \hat{\chi}^{-1}
\left(\hat{T}\hat{\chi}^{-1}\hat{T}\right)^{-1}
\hat{\chi}^{-1}\hat{L},
\label{oid}
\end{eqnarray}
where $\hat{T}=\hat{1}-\hat{L}$ is the transverse projector,
we can write for the inverse scalar response function
\begin{eqnarray}
\chi^{-1}\!=\!
-\frac{\omega^2}{c}
\nabla^{-2} \nabla \! \cdot \!
\left[
\hat{\chi}^{-1} \!\! - \!
\hat{\chi}^{-1} \! \left(\hat{T} \hat{\chi}^{-1}\hat{T}\right)^{\!-\!1} \!\!\!
\hat{\chi}^{-1}
\right]
\! \cdot \! \nabla \nabla^{-2}.
\label{chi1ts}
\end{eqnarray}
Then by Eq. (\ref{fs}) we have
\begin{eqnarray}
f_{xc}= -
\frac{\omega^2}{c}
\nabla^{-2} \nabla \cdot
\left\{
\hat{f}_{xc}
+\left(\hat{\chi}^{-1}_{KS} \!-\! \hat{f}_{xc}\right)\! \times
\right. \cr\cr
\!\!\!\left.
\left[\hat{T}\left(\hat{\chi}^{-1}_{KS} \!-\! \hat{f}_{xc}\right)\! \hat{T} \right]^{-1}\!\!\!
\left(\hat{\chi}^{-1}_{KS} \!-\! \hat{f}_{xc}\right)\!-\! \hat{\chi}^{-1}_{KS}
\left(\hat{T}\hat{\chi}^{-1}_{KS}\hat{T}\right)^{-1}\!\!\!
\hat{\chi}^{-1}_{KS}
\right\}
\! \cdot \! \nabla \nabla^{-2}.
\label{via}
\end{eqnarray}

Equation (\ref{via}) constitutes a formal expression for the scalar xc kernel $f_{xc}$
of the ordinary TDDFT in terms of the tensorial xc kernel $\hat{f}_{xc}$ of the TDCDFT and
the independent-particle (KS) tensorial response function $\hat{\chi}_{KS}$.
Then by virtue of Eqs.~(\ref{via}), (\ref{Q2}), (\ref{ft}), and (\ref{LP}) one can write
\begin{eqnarray}
&&Q_2=-
\frac{1}{c\,v^2}\,
\lim_{\omega\rightarrow 0} \omega\, {\rm Im}\,
(n_0-\bar{n}_0) {\bf v} \cdot
\hat{L} \left\{ \hat{f}_{xc}
+  \left(\hat{\chi}^{-1}_{KS} \!-\! \hat{f}_{xc}\right)\!
\left[\hat{T}  \left(\hat{\chi}^{-1}_{KS}\!-\!\hat{f}_{xc}\right)\! \hat{T}\right]^{-1}
\right. \cr\cr
&&\times \left. \left(\hat{\chi}^{-1}_{KS}-\hat{f}_{xc}\right)
-
\hat{\chi}^{-1}_{KS}
\left(\hat{T}\hat{\chi}^{-1}_{KS}\hat{T}\right)^{-1}
\hat{\chi}^{-1}_{KS}
\right\} \hat{L}
 (n_0-\bar{n}_0) {\bf v}.
\nonumber
\end{eqnarray}
In the above equation, the operator in the braces
is longitudinal from both sides, the fact of which can be verified by
applying the $\hat{T}$ operator from either side yielding
zero identically. We therefore can drop $\hat{L}$ operators to the result
\begin{eqnarray}
&&Q_2=-
\frac{1}{c\,v^2}\,
\lim_{\omega\rightarrow 0} \omega\, {\rm Im}\,
(n_0-\bar{n}_0) {\bf v} \cdot\
\left\{ \hat{f}_{xc}
+  \left(\hat{\chi}^{-1}_{KS} \!-\! \hat{f}_{xc}\right)\!
\left[\hat{T}  \left(\hat{\chi}^{-1}_{KS}\!-\!\hat{f}_{xc}\right)\! \hat{T}\right]^{-1}
\right. \cr\cr
&&\times \left. \left(\hat{\chi}^{-1}_{KS}-\hat{f}_{xc}\right)
-
\hat{\chi}^{-1}_{KS}
\left(\hat{T}\hat{\chi}^{-1}_{KS}\hat{T}\right)^{-1}
\hat{\chi}^{-1}_{KS}
\right\}
\cdot (n_0-\bar{n}_0) {\bf v}.
\label{Q2lim}
\end{eqnarray}

Equation (\ref{Q2lim}) is a formal expression
for the dynamical part of the friction coefficient in terms of the TDCDFT quantities.
It, however, is not easy to use in calculations. We, therefore, proceed
to derive an equation which will be shown in Sec.~\ref{Num} convenient to implement in practical calculations.
First, we introduce the notations
\begin{eqnarray}
&&{\bf x}=\left(\hat{T}\hat{\chi}^{-1}_{KS}\hat{T}\right)^{-1}
\hat{\chi}^{-1}_{KS}
\cdot (n_0-\bar{n}_0) {\bf v},
\label{xdf} \\
&&{\bf y}= \left(\hat{T} \hat{\chi}^{-1} \hat{T}\right)^{-1}
\hat{\chi}^{-1} \cdot (n_0-\bar{n}_0) {\bf v}.
 \label{df}
\end{eqnarray}
Then this is straightforward to show that ${\bf y}$ satisfies the equation
\begin{eqnarray}
{\bf y}= {\bf x}+\left(\hat{T} \hat{\chi}^{-1}_{KS} \hat{T}\right)^{-1}
\hat{f}_{xc}\cdot \left[{\bf y}-(n_0-\bar{n}_0) {\bf v}\right].
\label{yy}
\end{eqnarray}
Finally, with use of Eq.~(\ref{yy}),  Eq.~(\ref{Q2lim}) can be rewritten as
\begin{eqnarray}
Q_2=-
\frac{1}{c\,v^2}\,
\lim_{\omega\rightarrow 0} \omega\,
{\rm Im}\,
[{\bf x}-(n_0-\bar{n}_0) {\bf v}] \cdot
 \hat{f}_{xc} \cdot [{\bf y}-(n_0-\bar{n}_0) {\bf v}].
\label{Q2Q2}
\end{eqnarray}

\subsection{The Local Density Approximation for the tensorial xc kernel}
\label{LDAtk}

Within the LDA to the TDCDFT one can write \cite{Vignale-97}
\begin{eqnarray}
&&\int
f_{xc,ij}({\bf r},{\bf r}',\omega)\,
s_j({\bf r}') d {\bf r}' = \frac{i c}{\omega} \times \cr\cr
&&\left[ - \nabla_i V_{xc}^{ALDA}({\bf r},\omega)+
\frac{1}{n_0({\bf r})} \nabla_j \, \sigma_{xc,ij}({\bf r},\omega) \right],
\label{G1}
\end{eqnarray}
where
\begin{eqnarray}
V_{xc}^{ALDA}({\bf r},\omega)=\frac{1}{i \omega} \, \epsilon''_{xc}[n_0({\bf r})] \, \nabla_j \, s_j({\bf r}),
\label{V_xcstat}
\end{eqnarray}
$\epsilon_{xc}(n)$ is the xc energy density of the homogeneous EG of density $n$,
\begin{eqnarray}
&&\sigma_{xc,ij}({\bf r},\omega)=
\tilde{\eta}_{xc}[n_0({\bf r}),\omega] \left[\nabla_j\, u_i({\bf r}) + \nabla_i\, u_j({\bf r})\right.\cr\cr
&&\left. -\frac{2}{3} \, \nabla_k u_k({\bf r}) \, \delta_{ij} \right]
+ \tilde{\zeta}_{xc}[n_0({\bf r}),\omega] \, \nabla_k u_k({\bf r}) \, \delta_{ij},
\label{sig}
\end{eqnarray}
where
\begin{eqnarray*}
{\bf u}({\bf r}) ={\bf s}({\bf r})/n_0({\bf r}),
\end{eqnarray*}
the viscosity coefficients are given by
\begin{eqnarray}
&&\tilde{\zeta}_{xc}(n,\omega)=-\frac{n^2}{i\omega}
\left[f_{xc,L}^h(n,\omega) -\frac{4}{3}f_{xc,T}^h(n,\omega) - \epsilon''_{xc}(n) \right],\cr\cr
&&\tilde{\eta}_{xc}(n,\omega)=-\frac{n^2}{i\omega} f_{xc,T}^h(n,\omega),
\label{zeet0}
\end{eqnarray}
and $f_{xc,L}^h(n,\omega)$ and $f_{xc,T}^h(n,\omega)$ are, respectively, the longitudinal and transverse
xc kernels of the homogeneous EG with the density $n$.

\subsection{Resolution within the TDCDFT of the problem of the finite friction
coefficient of vacuum }
\label{RVP}
To show that within the LDA to the TDCDFT the friction coefficient of vacuum
is zero, it is sufficient to prove that with the tensorial xc kernel of the previous
subsection and the scalar xc kernel obtained from it by Eq.~(\ref{via}),
the sum rule of Eq.~(\ref{srV}) holds. We proceed by recalling exact sum rules for
the tensorial quantities \cite{Vignale-B}
\begin{eqnarray}
&&\int
\hat{f}_{xc,ij}({\bf r},{\bf r}',\omega) \, n_0({\bf r}') d {\bf r}'=
-\frac{c}{\omega^2} \nabla_i \nabla_j V_{xc}({\bf r}),
\label{sumbrf}\\
&&\frac{c}{\omega^2} \int
\hat{\chi}_{KS,ik}({\bf r},{\bf r}',\omega) \, \nabla'_k \nabla'_j V_{KS}( {\bf r}')
\, d {\bf r}' =
c \int
\hat{\chi}_{KS,ij}({\bf r},{\bf r}',\omega) \, d {\bf r}' - n_0({\bf r})\,\delta_{ij},
\label{sumbrKS}\\
&&\frac{c}{\omega^2} \int
\hat{\chi}_{ik}({\bf r},{\bf r}',\omega) \, \nabla'_k \nabla'_j V_0( {\bf r}')
\, d {\bf r}' =
c \int
\hat{\chi}_{ij}({\bf r},{\bf r}',\omega) \, d {\bf r}' - n_0({\bf r})\,\delta_{ij},
\label{sumbr0}
\end{eqnarray}
where $V_{xc}({\bf r})$ is the static xc potential.
With the LDA  to $\hat{f}_{xc}$ of the previous section, the sum rules
(\ref{sumbrf}) - (\ref{sumbr0}) are satisfied by construction
\cite{Vignale-B}.
In  Appendix~\ref{srVproof} we prove that {\it for a finite system}
the tensorial sum rules (\ref{sumbrf}) - (\ref{sumbr0})
entail the scalar sum rule (\ref{srV}).
As soon as the latter is satisfied,  results of Sec.~\ref{LDAS} lead
to the zero friction coefficient of vacuum within the LDA to the TDCDFT.

\section{Calculational procedures, results, and discussion}
\label{Num}
\begin{figure}[h]
\includegraphics{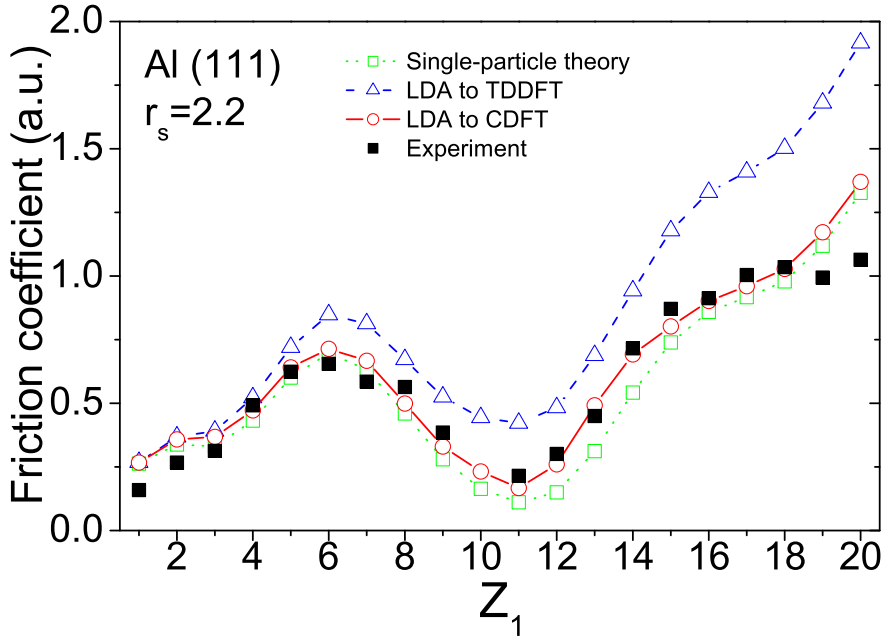}
\caption{\label{Fig_Al}
Friction coefficient of a homogeneous EG of $r_s=2.2$
versus the projectile
charge $Z_1$.
The open squares are the results of the calculation with  neglect of the dynamical xc as obtained from
Eq.~(\ref{start1}).
Triangles are the results of the calculation with the dynamical xc included
within the LDA to the conventional TDDFT as obtained from Eq.~(\ref{fxcl}).
Open circles are the results of the calculation
with the dynamical xc included within the LDA to the TDCDFT as described in Sec.~\ref{Num}.
Solid squares  are the measured SP of Al of
Ref.~\protect\cite{Winter-03} for slow ions ($v=0.5~{\rm a.u.}$) moving
at a distance of 1.2  a.u. from the last atomic plane of the Al (111) surface.}
\end{figure}
\begin{figure}[h]
\includegraphics{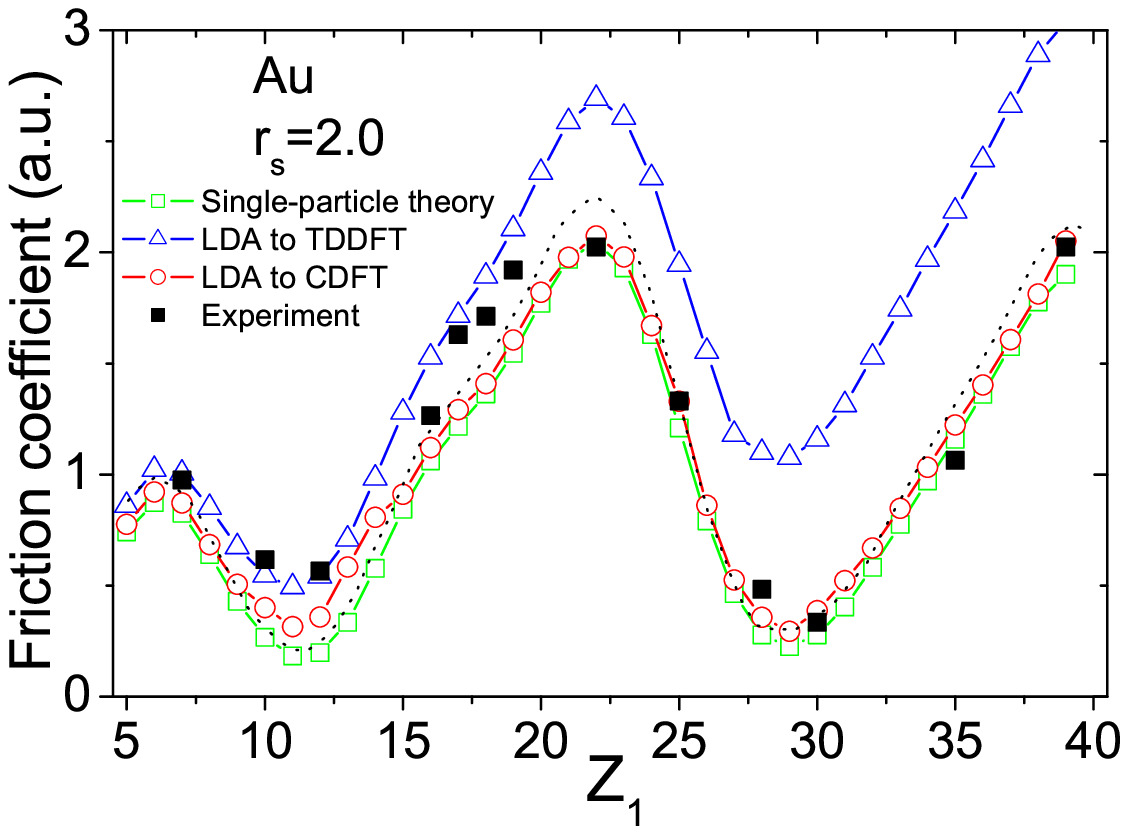}
\caption{\label{Fig_Au} Friction coefficient of a homogeneous EG of $r_s=2$
versus the projectile
charge $Z_1$.
Open squares are the results of the calculation with  neglect of the dynamical xc as obtained from
Eq.~(\ref{start1}).
Triangles are the results of the calculation with the dynamical xc included
within the LDA to the conventional TDDFT as obtained from Eq.~(\ref{fxcl}).
Open circles are the results of the calculation
with the dynamical xc included within the LDA to the TDCDFT as described in Sec.~\ref{Num}. Solid squares are the measurements
of Ref.~\protect\cite{Bottiger-69} of the SP of Au for slow ions
($v=0.68~{\rm a.u.}$) channeled along the (110) direction.
The dotted line is the calculation of Ref.~\protect\cite{Nagy2-89}
with the dynamical xc effects included in the framework
of the linear-response theory.}
\end{figure}
\begin{figure}[h]
\includegraphics{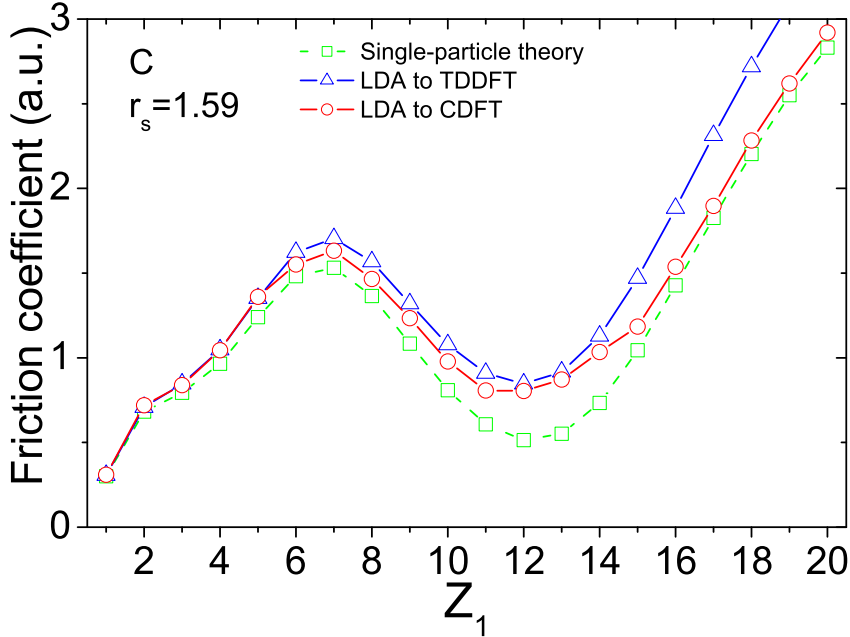}
\caption{\label{Fig_C} Friction coefficient of a homogeneous EG of carbon density ($r_s=1.59$)
versus the projectile
charge $Z_1$.
Open squares are the results of the calculation with neglect of the dynamical xc as obtained from
Eq.~(\ref{start1}).
Open triangles are the results of the calculation with the dynamical xc included
within the LDA to the conventional TDDFT as obtained from Eq.~(\ref{fxcl}).
Open circles are the results of the calculation
with the dynamical xc included within the LDA to the TDCDFT as described in Sec.~\ref{Num}.
Solid circles, triangles, and diamonds are
transmission measurements from Refs.~\protect\cite{Ormrod-6365},
\protect\cite{Ward-79s}, and \protect\cite{Hogberg-71} of the random SP of C
for ions moving with velocities $v=0.41$, $0.83$, and $0.25~{\rm a.u.}$, respectively.}
\end{figure}
We solve Eq.~(\ref{yy}) for the ${\bf y}$ vector at finite frequencies,
then substitute the results into Eq.~(\ref{Q2Q2}).
The friction
coefficient is found by the extrapolation from finite to zero frequency.
To conveniently treat $\left(\hat{T} \hat{\chi}^{-1}_{KS} \hat{T}\right)^{-1}$ in
the right-hand side of Eq.~(\ref{yy}), we use the identity
\begin{eqnarray}
\left(\hat{T} \hat{\chi}^{-1}_{KS} \hat{T}\right)^{-1}
%=
%\hat{\chi}_{KS}-\hat{\chi}_{KS} \left(\hat{L} \hat{\chi}_{KS} \hat{L}\right)^{-1} \hat{\chi}_{KS}\cr\cr
=
\hat{\chi}_{KS}+\frac{c}{\omega^2}
\hat{\chi}_{KS} \cdot {\bf \nabla} \chi^{-1}_{KS} {\bf \nabla} \cdot \hat{\chi}_{KS},
\label{id}
\end{eqnarray}
and equations (\ref{chifirst})-(\ref{Gfeq2}) and (\ref{fxcfirst}) - (\ref{fxclast})
are utilized for the explicit evaluation of the $\hat{\chi}_{KS}$ and $\hat{f}_{xc}$ operators, respectively.
With these provisions, the numerical procedure  is
to approximate Eq.~(\ref{yy}) by using a complete ortho-normal set of basis functions
thus reducing the problem to a system of linear equations.
The natural choice for the radial basis functions is
\begin{eqnarray}
\phi_n(r)=\left[\alpha^3 (n+1) (n+2)\right]^{-1/2} \, e^{-r/(2 \alpha)} \, L^{(2)}_n
(r/\alpha), \label{ba}
\end{eqnarray}
where $L^{(k)}_n(x)$ are the associated Laguerre polynomials
\footnote{$k=2$ ensures the orthonormality of the radial basis functions
with the weight of $r^2$.},
and $\alpha$ is a scaling parameter. Since the exact results
should not depend on the choice of the basis functions, we have been checking the convergence
of the numerical procedure by varying $\alpha$ and making sure that the respective results differ
insignificantly.

In Fig.~\ref{Fig_Al}, we plot the
friction coefficient of EG of $r_s=2.2$ versus the atomic number of a moving ion.
Results of the calculations
with neglect of the dynamical xc [the binary-collisions approximation, Eq.~(\ref{start1})],
the LDA to TDDFT [the sum of Eq.~(\ref{start1}) and Eq.~(\ref{fxcl})], and the LDA to TDCDFT
[the sum of Eq.~(\ref{start1}) and Eq.~(\ref{Q2Q2})] are shown together with the
experimental data of Ref. \cite{Winter-03} for ions moving with the velocity
of 0.5 a.u. at the distance of 1.2 a.u. from the last atomic plane of the (111) surface
of aluminum. The inhomogeneity of the electron density the ions travel through is not
strong under these conditions, and in the calculations we have used $r_s$ estimated
experimentally \cite{Winter-03}. Moreover, the experimental SP is
predominantly electronic since the trajectory of the projectile ion
remains well separated from the lattice atoms. Together, these two conditions
justify the comparison with the theory within the homogeneous EG model.
The non-monotonic dependence of the friction coefficient on the atomic number of the projectile
(the so-called $Z_1$-oscillations) is known to result within the single-particle theory
from the competition between
the increase in the EG-ion interaction with the growing charge of the bare nucleus
of the ion and the decrease of the same interaction due to the screening by the formation of
shells of bounded electrons of the pseudo-atom as well as its resonant states, as discussed in detail
in Ref. \cite{Echenique-86}.

While the LDA to the conventional TDDFT (triangles in Fig.~\ref{Fig_Al}) largely
overestimates the friction coefficient at $Z_1\ge5$,
the LDA to the TDCDFT (open circles in Fig.~\ref{Fig_Al}) is in a good agreement with the experiment
in a wide range of $3\le Z_1 \le 18$.
The significant deviation of our results from the experiment occurs
at $Z_1=1,2,19,20$, where the experimental friction coefficient
is {\em lower} than the binary-collisions calculations (open circles in Fig.~\ref{Fig_Al}).
This feature has been recently shown to be a single-particle effect due to the finite velocity
of the projectiles \cite{Vincent-07}, hence it is an effect of the deviation from the linear dependence
of the SP on the velocity. The same effect gives  {\em positive} contribution
at $8\le Z_1 \le 12$ suggesting that combined with the many-body effects
of the present theory the agreement with the experiment can be further improved.
\footnote{There is no reason, of course, for these two effects to be
additive.
It must be noted, that Ref. \cite{Vincent-07} incorrectly
attributes the overestimation by Ref. \cite{Nazarov-05}
of the contribution of the many-body effects
to the use of the {\em total} ground-state density
rather than the density of the delocalized states only.
The total particle-density is, however, the basic variable of the TDDFT
and without any artificial assumptions it enters the rigorous result of
Eq. (\ref{Q2}). The real source of the overestimation of the dynamical xc effects
in Ref. \cite{Nazarov-05} was, as Ref. \cite{Nazarov-05} had anticipated
and the present work shows, use of the LDA within the conventional TDDFT. The present work
is overcoming this shortcoming  within the LDA to the TDCDFT.
}
In the range $13\le Z_1 \le 17$, the dynamical many-body effects seem to be solely
responsible for the enhancement of the friction coefficient compared with
the binary-collisions calculation results.

In Fig.~\ref{Fig_Au}, we plot the friction coefficient
of EG of $r_s=2$ versus the atomic number of ions in the
range $5\le Z_1 \le 39$. This is compared with the available measured SP
for ions with the velocity of $0.68~{\rm a.u.}$ channeled along the (110) direction in gold.
Now because of the channeling, the collisions with the lattice atoms again do not give significant
contribution to the SP. We can see that the general trend is that the LDA to TDCDFT
improves the agreement between the theory and the experiment.
However, within the range $16 \le Z_1 \le 19$ the dynamical xc contribution
is too small to account for the onset at the experimental data,
nor can the persistent enhancement of the friction coefficient in this range be attributed to the
finite velocity effects within the single-particles theory \cite{Vincent-07}.
Further studies are required to elucidate the nature of this onset,
one of the possible causes being evidently the band-structure effects.

In Fig.~\ref{Fig_C}, we plot the results for EG of $r_s=1.59$
corresponding to the valence electron-density of carbon. It is instructive
to note that within the range $1 \le Z_1 \le 14$ the results within the LDA to the TDCDFT and
the LDA to the TDDFT do not differ significantly, which can be believed to be true in the general
case for light atoms in the high-density EG. Then at higher $Z_1$ rather abruptly the dynamical
xc contribution almost vanishes, which effect also can be understood qualitatively
recalling that for heavy atoms immersed in EG the charge-density distribution
(the main quantity entering the theory within the LDA to the TDCDFT)
is close to that of isolated atoms, and hence according to the results
of Sec.~\ref{RVP} it should disappear.
The experimental data in Fig.~\ref{Fig_C} correspond to the SP for ions moving along random
trajectories inside bulk carbon and, therefore, they are strongly influenced
by the collisions with lattice atoms, which fact hinders the quantitative
comparison with the calculations within the EG model.

It must be noted that the LDA to TDCDFT has no strict justification in
the $\omega\rightarrow 0$ limit.\cite{Vignale-96}
Nevertheless, the calculations of the {\em static} polarizabilities
of conjugated polymers using the Vignale-Kohn functional
in the form of Eqs.~(\ref{V_xcstat})-(\ref{zeet0}) have proved to be very successful.\cite{Faasen-02,Faasen-03}
In this respect, the situation seems to be analogous to that with the LDA to static DFT,
which had been justified for weakly varying electron densities only, while
the range of its successful applications has proved to be incomparably broader.
Similar to the method of Refs.~\cite{Faasen-02,Faasen-03}, we have been performing our calculations
at finite $\omega$ then extrapolating to $\omega=0$.

\section{Conclusions}
\label{Conc}
Within the current-density functional theory,
we have derived equations for the dynamical contribution to the
friction coefficient for
ions moving in electron gas. The resulting theory demonstrates a crucial
advantage over the ordinary time-dependent density-functional
approach since within the local-density approximation
the former ensures the correct zero friction coefficient in the limit of zero density of the electron gas,
while the latter does not.

We have traced the breakdown
of the LDA to the ordinary TDDFT
to the violation of a previously known sum rule
for the scalar exchange and correlation kernel.

Finally, we have performed calculations within the framework of the new theory
with the results finding themselves in a good overall agreement with the experimental
electronic stopping power of (i) the near-surface region of aluminum for slow ions moving in the grazing geometry
and (ii) gold for slow ions moving under  a channeling condition.

\appendix{Proof of Eq.~(\ref{Qint})}
\label{P1}
For a projectile moving in a homogeneous medium,
we can rewrite Eq.~(\ref{0}) in the reciprocal space as
\begin{eqnarray*}
-\frac{d E}{d x}= (2\pi)^3 \, {\rm Im}
\int \frac{{\bf v}\cdot{\bf q}}{v} \, v_q \, n_{ind}(q) \, d q,
\end{eqnarray*}
where $q=({\bf q},\omega)$ and
\begin{eqnarray*}
v_q=\frac{Z_1}{2\pi^2 {\bf q}^2}.
\end{eqnarray*}
An external potential $\phi_{ext}(q)$ induces in the homogeneous EG the particle density
\begin{eqnarray*}
n_{ind}(q) =
\sum\limits_{n=1}^{\infty} \int d q_1 ... d q_{n-1}
\chi_n(q,q_1,...,q_{n-1}) \phi_{ext}(q-q_1)  \cr\cr
\times \phi_{ext}(q_1-q_2) ... \phi_{ext}(q_{n-2}-q_{n-1}) \phi_{ext}(q_{n-1}),
\end{eqnarray*}
where $\chi_n(q,q_1,...q_{n-1})$
is the $n$-fold many-body density-response function of the homogeneous EG which satisfies
$n-1$ symmetry relations
\begin{eqnarray}
\chi_n(q_0,...q_j,q_{j+1},...,q_{n-\!1})\!=\! \cr\cr
\chi_n(q_0,...q_j,q_j\!-\!q_{n-\!1},...,q_j\!-\!q_{j\!+1\!}), \
0\le j < n-1.
\label{sym}
\end{eqnarray}

Since the bare potential by the projectile is
\begin{eqnarray*}
\phi_{ext}(q)=v_q \, \delta(\omega-{\bf v}\cdot {\bf q} ),
\end{eqnarray*}
we can write
\begin{eqnarray*}
n_{ind}(q)=\delta(\omega-{\bf v}\cdot {\bf q} ) \sum\limits_{n=1}^\infty
\int
\chi_n(q,q_1,...,q_{n-1})\cr\cr
\times v_{q_{n-1}} v_{q_{n-2}-q_{n-1}} ...v_{q-q_1} \, d {\bf q}_1 ... d {\bf q}_{n-1},
\end{eqnarray*}
where now $q=({\bf q},{\bf v}\cdot {\bf q})$.
Then we can write for the stopping power
\begin{eqnarray}
-\frac{d E}{d x}= (2 \pi)^3 \,{\rm Im} \sum\limits_{n=1}^\infty  \int
 \frac{{\bf v}\cdot{\bf q}_0}{v} \cr\cr
\times f_n(q,q_1,...,q_{n-1})
\, d {\bf q}_0 ... d {\bf q}_{n-1},
\label{st}
\end{eqnarray}
where we have introduced the notation
\begin{eqnarray}
f_n(q_0,q_1,...,q_{n-1})=
\chi_n(q_0,q_1,...,q_{n-1})\cr\cr
\times v_{q_{n-1}} v_{q_{n-2}-q_{n-1}} ...v_{q_0-q_1} v_{q_0}.
\label{f}
\end{eqnarray}
For the friction coefficient, which is the stopping power divided by the projectile velocity at
its zero value, we can write
\begin{eqnarray}
Q &=&(2 \pi)^3 \,{\rm Im}
\sum\limits_{n=1}^\infty
\sum\limits_{j=0}^{n-1}
\int
\frac{({\bf v}\cdot{\bf q}_0)}{v}
\frac{({\bf v}\cdot{\bf q}_j)}{v} \cr\cr
&\times& f_{nj}({\bf q}_0,...,{\bf q}_{n-1})
d {\bf q}_0...d {\bf q}_{n-1},
\label{Q}
\end{eqnarray}
where we have introduced further notations
\begin{eqnarray*}
f_{nj}({\bf q}_0,...,{\bf q}_{n-1})\!=\! \frac{\partial f_n(q_0,...,q_{n-\!1})}{\partial \omega_j}
\Bigr |_{\, \omega_0=0,...,\omega_{n-\!1}=0}.
\end{eqnarray*}

Our purpose is to prove that
\begin{equation}
Q=P,
\label{QP}
\end{equation}
where
\begin{eqnarray}
P  =
(2\pi)^3 \, {\rm Im} \int  d{\bf q} \, d{\bf k}
\frac{({\bf q}\cdot {\bf v})}{v}
\frac{({\bf k}\cdot {\bf v})}{v}
v_q v_k  \cr\cr \times \frac{\partial \chi({\bf q},{\bf k},\omega)}{\partial\omega}
\Bigr |_{\omega=0},
\label{P}
\end{eqnarray}
and $\chi({\bf q},{\bf k},\omega)$ is the linear density-response function of the system of the charge $Z_1$
at rest in the EG at origin.
The density induced
in a homogeneous EG by a wave-vector and frequency dependent
external perturbation $\psi_{ext}({\bf q},\omega)$ plus
the static potential of the charge is
\begin{eqnarray}
&&n_{ind}(q)=\sum\limits_{n=1}^\infty \int d q_1 ... d q_{n-1}
\chi_n(q, q_1, ... q_{n-1})\cr\cr
&&\times \phi_{ext}(q-q_1) ...\phi_{ext}(q_{n-2}- q_{n-1}) \phi_{ext}(q_{n-1}),
\label{n}\\
&&\phi_{ext}(q) = \psi_{ext}(q)+v_q \, \delta(\omega).
\label{phi}
\end{eqnarray}
Hence, to express the linear response function of the combined
system of the EG with the charge $Z_1$ at the origin,
we must collect in Eq.~(\ref{n}) terms linear in
$\psi_{ext}(q)$. With the use of symmetries (\ref{sym}), we readily arrive
at the result
\begin{eqnarray}
\chi({\bf q},{\bf k},\omega)&=& \chi_1({\bf q},\omega) \delta({\bf q}-{\bf k}) +
2 v_{q-k} \chi_2({\bf q},\omega,{\bf k},\omega) \cr\cr
&+& \sum\limits_{n=3}^\infty n
\int \chi_n({\bf q},\omega,{\bf q}_1,\omega,...,{\bf q}_{n-2},\omega,{\bf k},\omega)
\cr\cr
&\times& v_{q- q_1} ...v_{q_{n-2}- k} d {\bf q}_1...d
{\bf q}_{n-2}, \nonumber
\end{eqnarray}
where the dependence on the wave-vectors and frequencies has been written explicitly.
Hence, by virtue of Eq.~(\ref{P})
\begin{eqnarray*}
P =
(2\pi)^3 \, {\rm Im}
\sum\limits_{n=1}^\infty n
\int
\frac{({\bf q}_0\cdot {\bf v})}{v}\,
\frac{({\bf q}_{n-1}\cdot {\bf v})}{v}\cr\cr
  \times \frac{\partial}{\partial\omega} \,
f_n({\bf q}_0,\omega,...,{\bf q}_{n-1},\omega)
d {\bf q}_0...d {\bf q}_{n-1}
\Bigr |_{\omega=0},
\end{eqnarray*}
or
\begin{eqnarray*}
P &=&(2\pi)^3 \, {\rm Im}
\sum\limits_{n=1}^\infty n
\sum\limits_{j=0}^{n-1}
\int
\frac{({\bf q}_0\cdot {\bf v})}{v}\,
\frac{({\bf q}_{n-1}\cdot {\bf v})}{v}\cr\cr
&\times& f_{nj}({\bf q}_0,...,{\bf q}_{n-1}) \,
d {\bf q}_0...d {\bf q}_{n-1}.
\end{eqnarray*}

From Eqs.~(\ref{sym}) and (\ref{f}) it follows that $f_n$ functions satisfy the same
symmetry relations as $\chi_n$
\begin{eqnarray}
f_n(q_0,...q_j,q_{j+1},...,q_{n-1})=\cr\cr
f_n(q_0,...q_j,q_j\!-\!q_{n-1},...,q_j\!-\!q_{j+1}),  0\le j < n\!-\!1.
\label{symf}
\end{eqnarray}

\subappendix{}
For brevity, throughout the derivations below by $\int ... $ we imply the integral
$\int d{\bf q}_1...d{\bf q}_{n-1} ...$.
By Eq.~(\ref{symf}), for $j<i$ we can write
\begin{eqnarray}
&&\int {\bf q}_j
f_{ni}({\bf q}_0,...,{\bf q}_{n-1}) =
-\int {\bf q}_j
f_{n,n-1}({\bf q}_0,...{\bf q}_{i-1},
{\bf q}_{i-1}- {\bf q}_{n-1},...,{\bf q}_{i-1}-{\bf q}_i)=\cr\cr
&&-\int {\bf q}_j
f_{n,n-1}({\bf q}_0,...{\bf q}_{i-1},
{\bf q}_{n-1},...,{\bf q}_{i+1},{\bf q}_i)=
-\int {\bf q}_j
f_{n,n-1}({\bf q}_0,...,{\bf q}_{n-1}),
\label{1}
\end{eqnarray}
where we have used the integration variables ${\bf q}_i,...,{\bf q}_{n-1}$ substitutions leaving
other variables intact. But for $j<n-1$
\begin{eqnarray}
\int {\bf q}_j
f_{n,n-1}({\bf q}_0,...,{\bf q}_{n-1}) &=&
-\int {\bf q}_j f_{n,n-1}({\bf q}_0,...,{\bf q}_{n-2},{\bf q}_{n-2}-{\bf q}_{n-1})\cr\cr
&=&-\int {\bf q}_j f_{n,n-1}({\bf q}_0,...,{\bf q}_{n-1}),
\end{eqnarray}
which shows that for $j<n-1$
\begin{eqnarray}
\int {\bf q}_j
f_{n,n-1}({\bf q}_0,...,{\bf q}_{n-1}) = 0,
\end{eqnarray}
and then by Eq.~(\ref{1})
\begin{eqnarray}
\int {\bf q}_j
f_{ni}({\bf q}_0,...,{\bf q}_{n-1}) =0, \ \ j<i.
\label{2}
\end{eqnarray}

Further, if $0<j\le i \le n-1$, we can write
\begin{eqnarray}
&&\int \! \! {\bf q}_i f_{nj}({\bf q}_0,...{\bf q}_k,{\bf q}_{k+1},...,{\bf q}_{n-1}) \! = \!
- \! \! \! \int \! {\bf q}_i f_{n,n+k-j}({\bf q}_0,...{\bf q}_k,{\bf q}_k-{\bf q}_{n-1},...,{\bf q}_k-{\bf q}_{k+1})=\cr\cr
&&\int \!  ({\bf q}_i \! - \! {\bf q}_k) f_{n,n+k-j}({\bf q}_0,...{\bf q}_k,{\bf q}_{n-1},...,{\bf q}_{k+1}) \! \! = \! \! \!
\int \! ({\bf q}_{n+k-i} \! - \! {\bf q}_k) f_{n,n+k-j}({\bf q}_0,...,{\bf q}_{n-1}).
\nonumber
\end{eqnarray}
The first term in the last expression disappears due to Eq.~(\ref{2}), and we have
\begin{eqnarray}
\int {\bf q}_i f_{nj}({\bf q}_0,...,{\bf q}_{n-1}) =
\int {\bf q}_{n+k-i} f_{n,n+k-j}({\bf q}_0,...,{\bf q}_{n-1}).
\label{zz}
\end{eqnarray}
If $i>j$, then $n+k-i<n+k-j$, and  using Eq.~(\ref{2}) again we have
\begin{eqnarray}
\int \!\!{\bf q}_j
f_{ni}({\bf q}_0,...,{\bf q}_{n-1})  d{\bf q}_0...d{\bf q}_{n-1}\!=\!0, i\ne 0,  j\ne i.
\label{ii1}
\end{eqnarray}
Putting $i=j$ and $k=j-1$ in Eq.~(\ref{zz}), we obtain
\begin{eqnarray}
\int {\bf q}_j
f_{nj}({\bf q}_0,...,{\bf q}_{n-1}) =
\int {\bf q}_{n-1}
f_{n,n-1}({\bf q}_0,...,{\bf q}_{n-1}), \ \ j\ne 0.
\label{i2}
\end{eqnarray}

\subappendix{}
For $0<j\le n-1$ we can write
\begin{eqnarray}
&&\int \sum\limits_{s=j}^{n-1} {\bf q}_s\,
f_{n0}({\bf q}_0,...,{\bf q}_{n-1}) =
\int \sum\limits_{s=j}^{n-1} {\bf q}_s\,
f_{n0}({\bf q}_0,...,{\bf q}_{j-1},{\bf q}_{j-1}-{\bf q}_{n-1},...,{\bf q}_{j-1}-{\bf q}_j) =\cr\cr
&&\int [(n-j) {\bf q}_{j-1}-\sum\limits_{s=j}^{n-1} {\bf q}_s]\,
f_{n0}({\bf q}_0,...,{\bf q}_{j-1},{\bf q}_{n-1},...,{\bf q}_j) =
\int [(n-j) {\bf q}_{j-1}-\sum\limits_{s=j}^{n-1} {\bf q}_s]\,
f_{n0}({\bf q}_0,...,{\bf q}_{n-1}).
\nonumber
\end{eqnarray}
Then
\begin{eqnarray}
2 \int \sum\limits_{s=j}^{n-1} {\bf q}_s\,
f_{n0}({\bf q}_0,...,{\bf q}_{n-1}) \, d{\bf q}_0...d{\bf q}_{n-1}=\cr\cr
(n-j) \int {\bf q}_{j-1}
f_{n0}({\bf q}_0,...,{\bf q}_{n-1}) \, d{\bf q}_0...d{\bf q}_{n-1}.
\label{preind}
\end{eqnarray}
Using Eq.~(\ref{preind}), we prove that
\begin{eqnarray}
\int {\bf q}_j\,
f_{n0}({\bf q}_0,...,{\bf q}_{n-1}) =\cr\cr
(n-j) \int {\bf q}_{n-1}\,
f_{n0}({\bf q}_0,...,{\bf q}_{n-1}), \ \ j>0
\label{pr2}
\end{eqnarray}
by induction from $j=n-1$ to $j=1$.
For $0<i\le n-1$, we can write
\begin{eqnarray*}
&&\int {\bf q}_i f_{n0} ({\bf q}_0,...,{\bf q}_{n-1}) =
\int {\bf q}_i \sum\limits_{j=0}^{n-1} f_{nj} ({\bf q}_0,{\bf q}_0-{\bf q}_{n-1}...,{\bf q}_0-{\bf q}_1) =
\int ({\bf q}_0-{\bf q}_i) \sum\limits_{j=0}^{n-1} f_{nj} ({\bf q}_0,{\bf q}_{n-1}...,{\bf q}_1) \, =
\cr\cr
&&\int ({\bf q}_0-{\bf q}_{n-i}) \sum\limits_{j=0}^{n-1} f_{nj} ({\bf q}_0,{\bf q}_1...,{\bf q}_{n-1}) =
\int ({\bf q}_0-{\bf q}_{n-i}) f_{n0} ({\bf q}_0,{\bf q}_1...,{\bf q}_{n-1}) -
\int {\bf q}_{n-i} f_{n,n-i} ({\bf q}_0,{\bf q}_1...,{\bf q}_{n-1}),
\end{eqnarray*}
and then for $0<i\le n-1$
\begin{eqnarray}
\int ({\bf q}_i+{\bf q}_{n-i}-{\bf q}_0) f_{n0} ({\bf q}_0,...,{\bf q}_{n-1}) =
- \int {\bf q}_{n-i} f_{n,n-i} ({\bf q}_0,...,{\bf q}_{n-1}).
\label{iiii}
\end{eqnarray}

With use of Eqs.~(\ref{pr2}) and (\ref{i2}), Eq.~(\ref{iiii}) gives
\begin{eqnarray}
\int (n {\bf q}_{n-1}-{\bf q}_0) f_{n0} ({\bf q}_0,...,{\bf q}_{n-1}) =
- \int {\bf q}_{n-1} f_{n,n-1} ({\bf q}_0,...,{\bf q}_{n-1}).
\label{iiiii}
\end{eqnarray}

\subappendix{}
We write
\begin{eqnarray*}
&&P-Q=
\frac{(2\pi)^3}{v^2} \, {\rm Im}
\sum\limits_{n=1}^\infty
\sum\limits_{j=0}^{n-1}
\int
({\bf q}_0\cdot {\bf v})\,
\left[ (n {\bf q}_{n-1}-{\bf q}_j) \cdot {\bf v} \right]
f_{nj}({\bf q}_0,...,{\bf q}_{n-1}) \,
d {\bf q}_0...d {\bf q}_{n-1}= \cr\cr
&&\frac{(2\pi)^3}{v^2}  {\rm Im} \! \!
\sum\limits_{n=1}^\infty
\! \! \int \! \! ({\bf q}_0 \! \cdot \! {\bf v}) \! \! \left\{ \! \!
\left[ (n {\bf q}_{n-1} \! - \! {\bf q}_0) \cdot {\bf v} \right]
f_{n0}({\bf q}_0,...,{\bf q}_{n- \! 1}) \! + \! \!
\sum\limits_{j=1}^{n-1}
\left[ (n {\bf q}_{n- \! 1} \! - \! {\bf q}_j) \cdot {\bf v} \right]
f_{nj}({\bf q}_0,...,{\bf q}_{n- \! 1}) \!
\right\} \!  d {\bf q}_0...d {\bf q}_{n- \! 1}\cr\cr
&&=\frac{(2\pi)^3}{v^2} {\rm Im} \! \!
\sum\limits_{n=1}^\infty \!
\int ({\bf q}_0 \! \cdot \! {\bf v}) \! \left\{
\left[ (n {\bf q}_{n-1} \! - \! {\bf q}_0) \cdot {\bf v} \right]
f_{n0}({\bf q}_0,...,{\bf q}_{n-1}) \! + \!
({\bf q}_{n-1} \cdot {\bf v})
f_{n,n- \! 1}({\bf q}_0,...,{\bf q}_{n- \! 1})
\right\} d {\bf q}_0...d {\bf q}_{n- \! 1} \! =0,
\end{eqnarray*}
where the second equality is due to Eqs.~(\ref{ii1}) and (\ref{i2})
and the fourth equality is due to Eq.~(\ref{iiiii}).

\appendix{Proof of Eqs.~(\ref{Q12}) - (\ref{Q2})}
\label{Pr1}

We can rewrite Eq.~(\ref{Qint}) in the form
\begin{eqnarray}
Q  =  \int
\frac{[\nabla_{\bf r} V_0(r)\cdot {\bf v}]}{v}  \frac{[\nabla_{{\bf r}'} V_0(r')\cdot {\bf v}]}{v}
\chi({\bf r},{\bf r}'',0)
\frac{\partial \,{\rm Im} \chi^{-1}({\bf r}'',{\bf r}''',\omega)}{\partial\omega}
\Bigr |_{\omega=0}
\, \chi({\bf r}''',{\bf r}',0)
d{\bf r}  \, d{\bf r}' \, d{\bf r}''  \, d{\bf r}''',
\nonumber
\end{eqnarray}
or with use of the static sum rule (\ref{sumchi}) of the next section
\begin{eqnarray}
Q  =  \int
\frac{[\nabla_{\bf r} n_0(r)\cdot {\bf v}]}{v}  \frac{[\nabla_{{\bf r}'} n_0(r')\cdot {\bf v}]}{v}
\frac{\partial \,{\rm Im} \chi^{-1}({\bf r},{\bf r}',\omega)}{\partial\omega}
\Bigr |_{\omega=0}
\, d{\bf r}  \, d{\bf r}'.
\nonumber
\end{eqnarray}
Taking use of Eq.~(\ref{fs}), we have
\begin{eqnarray}
Q  =  \int
\frac{[\nabla_{\bf r} n_0(r)\cdot {\bf v}]}{v}  \frac{[\nabla_{{\bf r}'} n_0(r')\cdot {\bf v}]}{v}
\left[
\frac{\partial \,{\rm Im} \chi_{KS}^{-1}({\bf r},{\bf r}',\omega)}{\partial\omega}
\Bigr |_{\omega=0}
-\frac{\partial \,{\rm Im} f_{xc}({\bf r},{\bf r}',\omega)}{\partial\omega}
\Bigr |_{\omega=0}
 \right]
\, d{\bf r}  \, d{\bf r}',
\nonumber
\end{eqnarray}
or
\begin{eqnarray}
&&Q  =  -\int
\frac{[\nabla_{\bf r} n_0(r)\cdot {\bf v}]}{v}  \frac{[\nabla_{{\bf r}'} n_0(r')\cdot {\bf v}]}{v}
\frac{\partial \,{\rm Im} f_{xc}({\bf r},{\bf r}',\omega)}{\partial\omega}
\Bigr |_{\omega=0}
\, d{\bf r}  \, d{\bf r}' +\cr\cr
- &&\int
\frac{[\nabla_{\bf r} n_0(r)\cdot {\bf v}]}{v}  \frac{[\nabla_{{\bf r}'} n_0(r')\cdot {\bf v}]}{v}
\chi_{KS}^{-1}({\bf r},{\bf r}'',0)
\frac{\partial \,{\rm Im} \chi_{KS}({\bf r}'',{\bf r}''',\omega)}{\partial\omega}
\Bigr |_{\omega=0}
\, \chi_{KS}^{-1}({\bf r}''',{\bf r}',0)
d{\bf r}  \, d{\bf r}' \, d{\bf r}''  \, d{\bf r}'''.
\nonumber
\end{eqnarray}
Finally, using the static sum rule (\ref{chi0sum}) to simplify the second
term, we immediately arrive at Eqs.~(\ref{Q12}) - (\ref{Q2}).

\appendix{Static sum rules for the scalar response-functions}
\label{SSR}
A static shift of an ion by an infinitesimal vector $\Delta {\bf r}$ must result
in the same shift of the ground-state electron density.
The perturbation
corresponding to this shift is
\begin{eqnarray*}
\Delta V_0({\bf r})= \Delta {\bf r}\cdot \nabla V_0(r)
\end{eqnarray*}
and the change in the electron particle-density due to this shift is
\begin{eqnarray*}
\Delta n_0({\bf r})= \Delta {\bf r}\cdot \nabla n_0(r),
\end{eqnarray*}
which leads us to the static sum-rule
\begin{eqnarray}
\int \chi({\bf r},{\bf r}',0) \nabla' V_0(r')\, d {\bf r}' = \nabla n_0(r).
\label{sumchi}
\end{eqnarray}
The same sum rule evidently holds for the non-interacting KS density-response
function
\begin{eqnarray}
\int \chi_{KS}({\bf r},{\bf r}',0) \nabla' V_{KS}(r')\, d {\bf r}' = \nabla n_0(r).
\label{chi0sum}
\end{eqnarray}

Finally, inverting the above two equations and using
Eq.~(\ref{fs}), we arrive at the static sum-rule for the scalar xc kernel
\begin{eqnarray*}
\int f_{xc}({\bf r},{\bf r}',0) \nabla' n_{0}(r')\, d {\bf r}' =  \nabla V_{xc}(r).
\end{eqnarray*}

We note, that although the sum rules of this Appendix
follow as the static limit from the corresponding
dynamical sum rules of Ref. \cite{Vignale-95}, the former hold
for extended systems as well, while the latter are valid for bounded (finite) systems
only, fact of which necessitates the independent derivation of this Appendix.

\appendix{Proof of the equivalence between Eqs.~(\ref{Q1}) and (\ref{start1})}
\label{Q1start1}
The imaginary part of the KS response-function
can be written as
\begin{eqnarray}
&&{\rm Im} \, \chi_{KS}({\bf q},{\bf k},\omega) = -\frac{1}{(2 \pi)^2} \int
[f(\epsilon_s)-f(\epsilon_p)]\langle {\bf s}^-|\, e^{i {\bf q r}}\,| {\bf p}^+\rangle
\nonumber \\
&&\times \langle {\bf p}^+|\,e^{-i {\bf k r}}\,|{\bf s}^-\rangle
\delta(\omega-\epsilon_p+\epsilon_s) \,d {\bf p} \, d {\bf s},
\label{imchi0}
\end{eqnarray}
where $| {\bf p}^+\rangle$, $| {\bf s}^-\rangle$, and $\epsilon_p$ are the single-particles outgoing
scattering states, incoming states \cite{Taylor}, and the energies in the potential
$V_{KS}(r)$, respectively, and $f$ is the Fermi function.
Equation~(\ref{imchi0}) can be rewritten as
\begin{eqnarray*}
&&{\rm Im} \, \chi_{KS}({\bf q},{\bf k},\omega) =
-\frac{1}{(2 \pi)^2}  \int
[f(\epsilon_s)-f(\epsilon_s+\omega)] \times
\nonumber \\
 &&\langle {\bf s}^-|\, e^{i {\bf q r}}\,| {\bf p}^+\rangle \langle {\bf p}^+|\,e^{-i {\bf k r}}\,|{\bf s}^-\rangle
\delta(\omega-\epsilon_p+\epsilon_s)  \,d {\bf p} \, d {\bf s},
\end{eqnarray*}
and expanded at small $\omega$ to
\begin{eqnarray}
&&{\rm Im} \, \chi_{KS}({\bf q},{\bf k},\omega) =
-\frac{ \omega}{(2 \pi)^2}   \int
\delta(\epsilon_p-\epsilon_F) \, \delta(\epsilon_s-\epsilon_F)
\nonumber \\
 &&
\times \langle {\bf s}^-|\, e^{i {\bf q r}}\,| {\bf p}^+\rangle \langle {\bf p}^+|\,e^{-i {\bf k r}}\,|{\bf s}^-\rangle
  \,d {\bf p} \, d {\bf s}.
\end{eqnarray}

Now we can write by virtue of Eq.~(\ref{Q1})
\begin{eqnarray}
&&Q_1 =  \frac{2  \pi }{ v^2}
\! \! \int
\! \! \delta(\epsilon_p-\epsilon_F)  \delta(\epsilon_s-\epsilon_F) \,\langle {\bf s}^-|\,
[{\bf v } \cdot \nabla V_{KS}({\bf r})] \,| {\bf p}^+\rangle  \cr\cr
&& \times \langle {\bf p}^+|\,[{\bf v} \cdot \nabla V_{KS}({\bf r})]\,|{\bf s}^-\rangle
  \,d {\bf p} \, d {\bf s},\nonumber
\end{eqnarray}
where the square brackets in $[{\bf v}\cdot \nabla V_{KS}({\bf r})]$ denote that the gradient applies to
$V_{KS}({\bf r})$ only, leaving the wave-functions intact. We can further write
\begin{eqnarray}
&&\langle {\bf s}^-|\,[{\bf v} \cdot \nabla V_{KS}({\bf r})] \,| {\bf p}^+\rangle=
\langle {\bf s}^-|\,{\bf v } \cdot \nabla V_{KS}({\bf r}) \,| {\bf p}^+\rangle-
\langle {\bf s}^-| V_{KS}({\bf r}) {\bf v} \cdot \nabla | {\bf p}^+\rangle=\cr\cr
&&\langle {\bf s} + G_0^- V_{KS} {\bf s}^-|\,{\bf v } \cdot \nabla V_{KS}({\bf r}) \,| {\bf p}^+\rangle
-
\langle {\bf s}^-|\, V_{KS}({\bf r}) \,{\bf v} \cdot \nabla\,| {\bf p} + G_0^+ V_{KS} {\bf p}^+\rangle,
\label{A}
\end{eqnarray}
where we have used the Lippmann-Schwinger equations \cite{Taylor}
\begin{eqnarray}
{\bf p}^\pm\rangle = {\bf p}\rangle + G_0^\pm V_{KS}\, {\bf p}^\pm\rangle.
\nonumber
\end{eqnarray}
From Eq.~(\ref{A}) we have
\begin{eqnarray}
\langle {\bf s}^-|\,[{\bf v} \cdot \nabla V_{KS}({\bf r})] \,| {\bf p}^+\rangle=
i \, {\bf v} \cdot {\bf  s} \, \langle {\bf s}|\, V_{KS}\,| {\bf p}^+\rangle-
i \, {\bf v} \cdot {\bf  p} \, \langle {\bf s}^-|\, V_{KS}\,| {\bf p}\rangle,
\nonumber
\end{eqnarray}
and using formulas  for the transition-matrix \cite{Taylor}
\begin{eqnarray}
t({\bf s} \leftarrow {\bf p})=\langle {\bf s}|\, V_{KS}\,| {\bf p}^+\rangle = \langle {\bf s}^-|\, V_{KS}\,| {\bf p}\rangle,
\nonumber
\end{eqnarray}
we have
\begin{eqnarray}
\langle {\bf s}^-|\,[{\bf v} \cdot \nabla V_{KS}({\bf r})] \,| {\bf p}^+\rangle=
i \, {\bf v} \cdot ({\bf s}-{\bf p})\, t({\bf s} \leftarrow {\bf p}),
\nonumber
\end{eqnarray}
and similarly
\begin{eqnarray}
\langle {\bf p}^+|\,[{\bf v} \cdot \nabla V_{KS}({\bf r})] \,| {\bf s}^-\rangle=
-i \, {\bf v} \cdot ({\bf s}-{\bf p})\, t^*({\bf s} \leftarrow {\bf p}).
\nonumber
\end{eqnarray}

We can then write for the noninteracting-electrons part of the friction coefficient
\begin{eqnarray}
Q_1 =  \frac{2 \, \pi}{ v^2}
\int
\delta(\epsilon_p-\epsilon_F) \, \delta(\epsilon_s-\epsilon_F) \,
[{\bf v} \cdot ({\bf s}-{\bf p})]^2
|t({\bf s} \leftarrow {\bf p})|^2   \,d {\bf p} \, d {\bf s}.
\label{butone}
\end{eqnarray}
Recalling the expression of the differential cross-section through
the $T$-matrix element \cite{Taylor}
\begin{eqnarray}
\frac{d \sigma(\theta_{\hat{{\bf s} {\bf p}}})}{d \Omega_{\hat{{\bf s} {\bf p}}}}=
(2\pi)^4 |t({\bf s} \leftarrow {\bf p})|^2
\nonumber
\end{eqnarray}
and performing some integrations in Eq.~(\ref{butone})
explicitly, we arrive at Eq.~(\ref{start1}).

\appendix{A proof that for a finite system
the tensorial sum rules of Eqs.~(\ref{sumbrf})-(\ref{sumbr0}) lead to the scalar sum rule
of Eq.~(\ref{srV})}
\label{srVproof}
Equation (\ref{sumbr0}) can be rewritten  as
\begin{eqnarray}
\frac{c}{\omega^2} \int
\hat{\chi}_{ik}({\bf r},{\bf r}',\omega) \, \nabla'_k \nabla'_j V_0( {\bf r}')
\, d {\bf r}' =
c \int
\hat{\chi}_{ik}({\bf r},{\bf r}',\omega) \,\nabla'_k r'_j \, d {\bf r}' - n_0({\bf r})\,\delta_{ij}.
\label{AAAA}
\end{eqnarray}
The next step, which involves integration by parts, requires
the response function to vanish at infinity and, therefore, it applies to bounded systems only.
For the latter case we can write
 multiplying Eq.~(\ref{AAAA}) scalarly from the left by $\nabla$ and using  Eq.~(\ref{chit})
\begin{eqnarray}
\int
\chi({\bf r},{\bf r}',\omega) \left[ \nabla'_j V_0( {\bf r}') - \omega^2 r'_j \right]
\, d {\bf r}' =
\nabla_j n_0({\bf r}),
\nonumber
\end{eqnarray}
and after inverting
\begin{eqnarray}
\int
\chi^{-1}({\bf r},{\bf r}',\omega) \nabla_j n_0({\bf r}')
\, d {\bf r}' =
\nabla_j V_0( {\bf r}) - \omega^2 r_j .
\label{Achi}
\end{eqnarray}
A similar relation holds for $\chi_{KS}$
\begin{eqnarray}
\int
\chi_{KS}^{-1}({\bf r},{\bf r}',\omega) \nabla_j n_0({\bf r}')
\, d {\bf r}' =
\nabla_j V_{KS}( {\bf r}) - \omega^2 r_j .
\label{AchiKS}
\end{eqnarray}
Subtracting Eq.~(\ref{Achi}) from Eq.~(\ref{AchiKS}) and using the definition of
Eq.~(\ref{fs}), we immediately arrive at Eq.~(\ref{srV})
[compare with Ref.~\cite{Vignale-95}].

\appendix{Explicit expressions for the KS current-density response
function, the xc kernel, and the ${\bf x}$ vector of Eq.~(\ref{xdf}) for a system with spherical symmetry}
\label{Expl}
\subappendix{KS current-density response function}
The KS response function can be explicitly written as
\begin{eqnarray}
&&\hat{\chi}_{KS,ij}({\bf r},{\bf r}',\omega) =
\frac{1}{ c} \, n_0(r) \delta({\bf r}-{\bf r}') \, \delta_{ij} -\frac{1}{4 c} \times \cr\cr
&&\sum\limits_{\alpha \beta} \frac{f_\alpha-f_\beta}{\omega-\epsilon_\beta+\epsilon_\alpha+i\eta}
\left[\psi^*_\alpha({\bf r}) \nabla_i \psi_\beta({\bf r})- \psi_\beta({\bf r}) \nabla_i \psi^*_\alpha({\bf r})\right]\cr\cr
&&\times \left[ \psi^*_\beta({\bf r}') \nabla'_j \psi_\alpha({\bf r}')- \psi_\alpha({\bf r}') \nabla'_j \psi^*_\beta({\bf r}') \right],
\label{chiKS}
\end{eqnarray}
where $\psi_\alpha({\bf r})$ and $\epsilon_\alpha$ are the single-particle wave-function
and eigenenergy, respectively, in the state $\alpha$, and $f_\alpha$ is the occupation number of this state.
From the spherical symmetry of the problem
it is easy to conclude that both $\hat{\chi}_{KS}$ and $\hat{f}_{xc}$
leave invariant the subspace of the vectors of the form
\begin{eqnarray}
a(r) \, {\bf v} + b(r) \, ({\bf r}\cdot{\bf v}) \, {\bf r},
\nonumber
\end{eqnarray}
where $a(r)$ and $b(r)$ are arbitrary scalar functions of $r=|{\bf r}|$,
and, hence, both ${\bf x}$ and the solution ${\bf y}$ of Eq.~(\ref{yy})
are the vectors from the same subspace. Furthermore, by the definition
(\ref{df}), ${\bf y}$ is the transverse vector,
which property imposes the fulfillment of the relation
\begin{eqnarray}
a'(r)+4\,r\,b(r)+ r^2\, b'(r) =0,
\nonumber
\end{eqnarray}
and, therefore, Eq.~(\ref{yy}) effectively becomes an equation
with respect to one unknown scalar function of the radial coordinate.
With use of Eq.~(\ref{chiKS}), we arrive at the equalities
which are sufficient to evaluate $\hat{\chi_{KS}}$-dependent quantities
in Eq.~(\ref{yy})
\begin{eqnarray}
&&c \int \hat{\chi}_{KS,ij}({\bf r},{\bf r}',\omega) a(r',\omega) \delta_{jk} d{\bf r}'= \tilde{a}(r,\omega) \delta_{ik}
+\tilde{b}(r,\omega) r_i r_k,
\label{chifirst}\\
&&c \int \hat{\chi}_{KS,ij}({\bf r},{\bf r}',\omega) \nabla'_j \nabla'_k F(r') d {\bf r}' =
k(r,\omega)\, \delta_{ik} + m(r,\omega) r_i r_k,
\label{chirightsc1}
\end{eqnarray}
where
\begin{eqnarray}
&&\tilde{a}(r)= n_0(r) \, a(r) + q(r),\cr\cr
&&\tilde{b}(r,\omega)= -q(r)/r^2+p(r)/r^4,
\nonumber
\end{eqnarray}
and
\begin{eqnarray}
&&q(r)=
-\frac{n_0(r)}{r} \int\limits_\infty^r a(r') d r'-
\frac{\omega}{4\pi r^3} \sum\limits_{\alpha, l_\beta}
f_\alpha \left(\delta_{l_{\alpha}-1,l_\beta} \, l_\alpha^2 -\delta_{l_{\alpha},l_\beta-1} \, l_\beta^2\right)
y_\alpha(r) \cr\cr
&&\times \int\limits_0^\infty d r'
\left[G_{l_\beta,\epsilon_\alpha+\omega}(r,r')-G^*_{l_\beta,\epsilon_\alpha-\omega}(r,r')\right]
y_\alpha(r') \int\limits_\infty^{r'} a(r'') dr'' \cr\cr
&&+
\frac{1}{4 \pi   r^3} \sum\limits_{\alpha, l_\beta} f_\alpha \left(\delta_{l_{\alpha}-1,l_\beta} \, l_\alpha^3 +\delta_{l_{\alpha},l_\beta-1} \, l_\beta^3\right)
y_\alpha(r)
\int\limits_0^\infty
dr' \left[G_{l_\beta,\epsilon_\alpha+\omega}(r,r')+G^*_{l_\beta,\epsilon_\alpha-\omega}(r,r')\right] \frac{y_\alpha(r')}{r'} \cr\cr
&& \times \left[a(r')-\frac{1}{r'} \int\limits_\infty^{r'} a(r'') dr'' \right] ,
\label{Gfeq1} \\
&&p'(r)=2 r q(r)-r^2[a(r) n_0(r)]' +
\frac{\omega}{2 \pi } \sum\limits_{\alpha,  l_\beta} f_\alpha \left(\delta_{l_{\alpha}-1,l_\beta} \, l_\alpha^2 -\delta_{l_{\alpha},l_\beta-1} \, l_\beta^2\right)
y_\alpha(r) \cr\cr
&&\times \int\limits_0^\infty
dr' \left[G_{l_\beta,\epsilon_\alpha+\omega}(r,r')-G^*_{l_\beta,\epsilon_\alpha-\omega}(r,r')\right]
\frac{y_\alpha(r') }{r'}
\left[a(r')-\frac{1}{r'} \int\limits_\infty^{r'} a(r'') dr'' \right]
-
\frac{\omega^2}{2 \pi }  \sum\limits_{\alpha, l_\beta}  f_\alpha \cr\cr
&&\times  \! \left(\delta_{l_{\alpha}-1,l_\beta} \, l_\alpha \! + \! \delta_{l_{\alpha},l_\beta-1} \, l_\beta\right) \!
y_\alpha(r) \! \! \!
\int\limits_0^\infty \! dr' \!
\left[G_{l_\beta,\epsilon_\alpha+\omega}(r,r') \! + \! G^*_{l_\beta,\epsilon_\alpha-\omega}(r,r')\right]
y_\alpha(r')   \! \! \int\limits_\infty^{r'} \! \! a(r'') dr'',
\label{Gfeq2}
\end{eqnarray}
\begin{eqnarray}
&&k(r) \! = \!
-\frac{\omega}{4 \pi  r^3} \! \sum\limits_{\alpha,l_\beta} \!
f_\alpha \! \! \left(l_\alpha^2  \delta_{l_\beta,l_\alpha-1} \! - \! l_\beta^2  \delta_{l_\alpha,l_\beta-1} \right)
y_{l_\alpha,k_\alpha}(r) \!
\int\limits_0^\infty \!
\left[ G_{l_\beta,\epsilon_\alpha+\omega}(r,r') \! - \! G^*_{l_\beta,\epsilon_\alpha-\omega}(r,r') \right]
y_{l_\alpha,k_\alpha}(r') F'(r') d r',\cr\cr
&&\frac{[r^4 m(r)]'}{r^2} =- k'(r)
-\frac{\omega^2}{2 \pi  r^2}
\sum\limits_{l_\beta, \alpha }
f_\alpha (l_\alpha \delta_{l_\beta,l_\alpha-1}+l_\beta \delta_{l_\alpha,l_\beta-1})
y_{l_\alpha,k_\alpha}(r) \cr\cr
&&\times \int\limits_0^\infty
[G_{l_\beta,\epsilon_\alpha+\omega}(r,r') +G^*_{l_\beta,\epsilon_\alpha-\omega}(r,r')]
 y_{l_\alpha,k_\alpha}(r') F'(r') d r',
 \label{MMM}
\end{eqnarray}
where $y_{\alpha}(r)$
are the solutions of the radial Schr\"{o}dinger equation
\begin{eqnarray}
\left[\frac{d^2}{dr^2}-\frac{l_\alpha (l_\alpha+1)}{r^2}- 2 V_{KS}(r)+ 2 \epsilon_\alpha\right]
y_\alpha(r)=0,
\nonumber
\end{eqnarray}
and
\begin{eqnarray}
G_{l,\epsilon}(r,r')= \sum\limits_{k_\beta} \frac{y_{l,k_\beta}(r)
\, y_{l,k_\beta}(r')}{\epsilon+i \eta -\epsilon_\beta}
\nonumber
\end{eqnarray}
is the Green's function.

\subappendix{Exchange and correlation kernel}
In the case of spherical symmetry, we can write
\begin{eqnarray}
\int \hat{f}_{xc,ij}({\bf r},{\bf r}',\omega)
\left[ a(r',\omega) \delta_{jk} + b(r',\omega) r'_j r'_k \right] d{\bf r}'= \tilde{a}(r,\omega) \delta_{ik}
+\tilde{b}(r,\omega) r_i r_k,
\label{fxcfirst}
\end{eqnarray}
and, with use of the equations of Sec.~\ref{LDAtk}, we have
\newcommand {\imag}{i}
\begin{eqnarray}
&&\tilde{a}(r)=\frac{\imag c}{3 r\,{\omega }^2\,{n_0}(r)}
    \left\{ 3\,\left[  4\,\imag  \,r\,b(r)\,
          \epsilon_{xc} (r)\,{n_0}(r) +
         \imag \,\epsilon_{xc} (r)\,{n_0}(r)\,
          \left( a'(r) + r^2\,b'(r) \right)  +
         r\,\omega \,\tilde{\eta}_{xc} '(r)\,
          \left( r\,{b_1}(r) + a'_1(r) \right)  \right] \right.\cr\cr
          &&\left.
          + 3\,\omega \,\tilde{\zeta}_{xc} (r)\,
       \left[ 4\,r\,{b_1}(r) + a'_1(r) +
         r^2\,b'_1(r) \right]  +
      \omega \,\tilde{\eta}_{xc} (r)\,
       \left[ 7\,a'_1(r) +
         r\,\left( 10\,{b_1}(r) + r\,b'_1(r) +
            3\,a''_1(r) \right)  \right]  \right\} \cr\cr
&&\tilde{b}(r)=
\frac{\imag c }{3 r^3\,
    {\omega }^2\,{n_0}(r)}\,
    \left\{ 12\,\imag \,r^2\,b(r)\,
       {n_0}(r)\,\epsilon_{xc} '(r) +
       3\,\imag   \,r\,{n_0}(r)\,a'(r)\,
       \epsilon_{xc} '(r) +  3\,\imag  \,r^3\,
       {n_0}(r)\,b'(r)\,\epsilon_{xc} '(r) \right.\cr\cr
       &&\left.+
      12\,r^2\,\omega \,{b_1}(r)\,\tilde{\zeta}_{xc} '(r) +
      r^2\,\omega \,{b_1}(r)\,\tilde{\eta}_{xc} '(r) -
      3\,\omega \,\tilde{\zeta}_{xc} (r)\,a'_1(r) -
      \omega \,\tilde{\eta}_{xc} (r)\,a'_1(r) +
      3\,r\,\omega \,\tilde{\zeta}_{xc} '(r)\,a'_1(r) \right.\cr\cr
      &&\left. +
      r\,\omega \,\tilde{\eta}_{xc} '(r)\,a'_1(r) +
      15\,r^2\,\omega \,\tilde{\zeta}_{xc} (r)\,b'_1(r) +
      23\,r^2\,\omega \,\tilde{\eta}_{xc} (r)\,b'_1(r) +
      3\,r^3\,\omega \,\tilde{\zeta}_{xc} '(r)\,b'_1(r) +
      4\,r^3\,\omega \,\tilde{\eta}_{xc} '(r)\,b'_1(r) \right.\cr\cr
      &&\left. -
      3\,\imag \,\epsilon_{xc} (r)\,{n_0}(r)\,
       \left[ a'(r) - r\,
          \left( 5\,r\,b'(r) + a''(r) + r^2\,b''(r) \right)
             \right]  +
      3\,r\,\omega \,\tilde{\zeta}_{xc} (r)\,a''_1(r) +
      r\,\omega \,\tilde{\eta}_{xc} (r)\,a''_1(r) \right. \cr\cr
      &&\left. +
      r^3\,\omega \,\left[ 3\,\tilde{\zeta}_{xc} (r) + 4\,\tilde{\eta}_{xc} (r)
         \right] \,b''_1(r) \right\},
\label{fxclast}
\end{eqnarray}
where $a_1(r)=a(r)/n_0(r)$ and $b_1(r)=b(r)/n_0(r)$.

We are using the LDA to the tensorial xc kernel of Sec.~\ref{LDAtk}, with
the low-frequency expansion of $f_{xc,L}^h(n,\omega)$ and $f_{xc,T}^h(n,\omega)$
up to the first order in $\omega$. Using the equalities \cite{Qian-02}

\begin{eqnarray}
&&f_{xc,L}^h(n,0) -\frac{4}{3}f_{xc,T}^h(n,0) - \epsilon''_{xc}(n)=0,\cr\cr
&&\frac{\partial f_{xc,L}^h(n,\omega)}{\partial \omega}\Bigr |_{\omega=0} -
\frac{4}{3}\frac{\partial f_{xc,T}^h(n,\omega)}{\partial \omega}\Bigr |_{\omega=0} =0,
\nonumber
\end{eqnarray}
we have from Eqs.~(\ref{zeet0})
\begin{eqnarray}
&&\tilde{\zeta}_{xc}(n,\omega)=0,\cr\cr
&&\tilde{\eta}_{xc}(n,\omega)=\frac{i n^2}{\omega}
\left[  f_{xc,T}^h(n,0) +i \, \omega \frac{\partial {\rm Im}\, f_{xc,T}^h(n,\omega)}{\partial \omega}\Bigr |_{\omega=0} \right].
\label{zeet}
\end{eqnarray}
For $f_{xc,T}^h(n,0)$, we take use of its expression through the shear modulus
$\mu_{xc}$
\begin{eqnarray}
f_{xc,T}^h(n,0)=\frac{\mu_{xc}(n)}{n^2}.
\nonumber
\end{eqnarray}
For the local density such that $1 \le r_s \le 5$
we obtain $\mu_{xc}(n)$ by the interpolation between the values of
Ref. \cite{Qian-02}.
For $r_s < 1$, we use the high-density approximation
\cite{Conti-99}
\begin{eqnarray}
\mu_{xc}(n)=\frac{n k_F(n)}{10 \pi}.
\nonumber
\end{eqnarray}
For $\frac{\partial f_{xc,T}^h(n,\omega)}{\partial \omega}\Bigr |_{\omega=0}$,
we are using Eq.~(15) of Ref.~\cite{Qian-02}.

\subappendix{The ${\bf x}$ vector of Eq.~(\ref{xdf})}
Using the definition of Eq.~(\ref{xdf})
together with Eq.~(\ref{id}) we can write
\begin{eqnarray}
{\bf x}(\omega) = (n_0- \bar n_0) {\bf v} + \frac{c}{\omega^2}
\hat{\chi}_{KS}(\omega) \nabla \chi_{KS}^{-1}(\omega) \nabla (n_0 {\bf v}),
\nonumber
\end{eqnarray}
where we have explicitly written the frequency dependence of the functions. Expanding to the first order in $\omega$
\begin{eqnarray}
\chi_{KS}^{-1}(\omega) \nabla (n_0 {\bf v})=
\chi_{KS}^{-1}(0) \nabla (n_0 {\bf v})+
\omega
\frac{\partial \chi_{KS}^{-1}(\omega)}{\partial \omega}\Bigr |_{\omega=0} \nabla (n_0 {\bf v})
\nonumber
\end{eqnarray}
and using the static sum rule (\ref{sumchi}), we can write

\begin{eqnarray}
{\bf x}(\omega) = (n_0- \bar n_0) {\bf v} + \frac{c}{\omega^2}
\hat{\chi}_{KS}(\omega) \nabla ({\bf v} \cdot \nabla V_{KS})\cr\cr
-
\frac{c}{\omega} \hat{\chi}_{KS}(\omega)
\nabla \chi_{KS}^{-1}(0)
\left[\frac{\partial \chi_{KS}(\omega)}{\partial \omega}\right]_{\omega=0}
({\bf v} \cdot \nabla V_{KS}).
\label{xxx}
\end{eqnarray}
It must be noted that an expansion in $\omega$ of $\hat{\chi}_{KS}(\omega)$ applied to a gradient
of an $\omega$-independent function starts from the $\omega^2$ term and, therefore,
the second and the third terms in the above expression have leading terms of $\omega^0$
and $\omega^1$, respectively. To evaluate Eq.~(\ref{xxx}) to the first order in $\omega$
we derive and use the following equalities:
\begin{eqnarray}
\frac{c}{\omega^2}
\hat{\chi}_{KS}(\omega) \nabla ({\bf v} \cdot \nabla V_{KS}) =
a(r,\omega)\, {\bf v} + b(r,\omega) \, ({\bf v}\cdot{\bf r}) \, {\bf r},
\label{chiab}
\end{eqnarray}

\begin{eqnarray}
&&a(r,\omega)=\frac{9 h(r,\omega)-15 g(r,\omega)}{16 \pi}-n_0(r),\cr\cr
&&b(r,\omega)=\frac{45 g(r,\omega)-15 h(r,\omega)}{16 \pi r^2},
 \label{ab}
\end{eqnarray}

\begin{eqnarray}
\left[
\begin{array}{c}
  h(r,\omega) \\
  g(r,\omega)
\end{array}
\right] =
\left[
\begin{array}{c}
  h_0(r) \\
  g_0(r)
\end{array}
\right]+
\omega \left[
\begin{array}{c}
  h_1(r) \\
  g_1(r)
\end{array}
\right] + ... ,
\end{eqnarray}

\begin{eqnarray}
\left[
\begin{array}{c}
  h_0(r) \\
  g_0(r)
\end{array}
\right] =
\frac{2 R}{ \pi r^3} \!\!\sum\limits_{ l=0}^\infty
\left[
\begin{array}{c}
  H_l(k_F,k_F,r) \\
  G_l(k_F,k_F,r)
\end{array}
\right]
\cos[\delta_{l-1}(k_F)-\delta_l(k_F)],
\end{eqnarray}

\begin{eqnarray}
&&\left[
\begin{array}{c}
  h_1(r) \\
  g_1(r)
\end{array}
\right] =
\frac{R}{  \pi k_F r^3} \!\!\sum\limits_{ l=0}^\infty
\left[
\begin{array}{c}
  H_l(k_F,k_F,r) \\
  G_l(k_F,k_F,r)
\end{array}
\right]
\left\{\cos[\delta_{l-1}(k_F)-\delta_l(k_F)] [\delta'_{l-1}(k_F)+\delta'_l(k_F)]
- \right. \cr\cr
&&\left. \frac{2 \, l}{k_F}
\sin[\delta_{l-1}(k_F)-\delta_l(k_F)]\right\}
+
\left\{\left(\frac{\partial }{\partial k_\beta} -
\frac{\partial }{\partial k_\alpha}\right) \left[
\begin{array}{c}
  H_l(k_\alpha,k_\beta,r) \\
  G_l(k_\alpha,k_\beta,r)
\end{array}
\right] \right\}_{k_\alpha=k_\beta=k_F} \cr\cr
&&\times \sin[\delta_{l-1}(k_F)-\delta_l(k_F)],
\end{eqnarray}

\begin{eqnarray}
\left[
\begin{array}{c}
  H_l(k_\alpha,k_\beta,r) \\
  G_l(k_\alpha,k_\beta,r)
\end{array}
  \right]     &=&     \left[
\begin{array}{c}
  2/3 \\
  2/15
\end{array}
  \right]
l^2 y_{l,k_\alpha} (r) y_{l   -   1,k_\beta} (r) \cr\cr &+&  \left[
\begin{array}{c}
  1/3 \\
  1/5
\end{array}
  \right]
l \, r \, [y_{l  -   1,k_\beta} (r)
y'_{l,k_\alpha} (r)  -  y_{l,k_\alpha} (r)y'_{l- 1,k_\beta} (r)].
\label{lala}
\end{eqnarray}

Equations (\ref{chiab}) - (\ref{lala}) are enough to explicitly evaluate the second term
and to apply the first operator from the right in the third term of Eq.~(\ref{xxx}). We, however,
did not find a way to explicitly apply $\chi_{KS}^{-1}(0)$ in the third term and, therefore,
we invert $\chi_{KS}(0)$ on the set of the basis functions of Eq.~(\ref{ba}) and finally use Eqs.~(\ref{chirightsc1})
and (\ref{MMM}).

\section*{Acknowledgements}

G. V. and Y. T. acknowledge, respectively, financial support by the Department
of Energy grant DE-FG02-05ER46203 and a Grant-in-Aid for Scientific
Research in Priority Areas (No.17064004) of MEXT, Japan.

\end{document}